\documentclass[prl,twocolumn]{revtex4}
\usepackage{amsmath,amsfonts,amssymb}
\usepackage{graphicx}

\begin{document}

\title{Multi-Channel Interference in Resonance-Like Enhancement of High-Order Above-Threshold Ionization}
\author{Long Xu}
\author{Libin Fu}
\email{lbfu@gscaep.ac.cn}
\affiliation{Graduate School of China Academy of Engineering Physics, No. 10 Xibeiwang
East Road, Haidian District, Beijing, 100193, China}

\begin{abstract}
The intensity-dependent resonance-like enhancement phenomenon in high-order
above-threshold ionization spectrum is a typical quantum effect for atoms or
molecules in the intense laser field, which has not been well understood.
The calculations of TDSE are in remarkable agreement with the experimental
data, but they can not clarify the contributions of the bound states.
The semi-classical approach of strong field approximation, in which no excited
states are involved, can obtain the similar phenomenon, but the laser
intensities of enhanced regions predicted by SFA are inconsistent with the
results of TDSE.
In this letter, a new fully quantum model is established from TDSE with not any excited states.
Two types of enhanced structures, unimodal and multimodal structures, are found in the results of TDSE and model.
Besides, the calculations of model reproduce the key features of the results of TDSE.
It shows that the excited states are not the key factor in the resonance-like enhancements in our calculated system, since there are not any excited states in our model.
Based on the calculations of our model, we show that such resonance-like enhancements are caused by the constructive interference of different momentum transfer channels.
Last, the Fano-like lineshapes are also discussed for the features of multi-channel interference.
\end{abstract}

\maketitle


%

Above-threshold ionization (ATI) is the most fundamental and essential
nonperturbative phenomenon in the interaction of the intense laser pulse
with atoms or molecules, and it has attracted extensive attention
\cite{Becker2012, Pazourek2015, Peng2015} due to various possible
applications since it was first observed forty years ago \cite{Agostini1979}.
ATI reveals that the electron can absorb more photons than required to
exceed the ionization threshold for atoms or molecules in the intense laser field.
The ATI spectra have been well understood by combining the
contributions of the direct ionization and the rescattering process.
For the electron kinetic energy less than 2$U_{p}$, the ATI spectrum decreases exponentially,
where 2$U_{p}$ is the maximal energy of a directly ionized electron
\cite{Paulus1994, Paulus1994_2, Becker2002}
and $U_{p}=E_{0}^{2}/(4\omega ^{2})$ is the ponderomotive energy.
$E_{0}$ and $ \omega $ are the laser peak intensity and laser frequency, respectively.
The high-energy spectrum appears a flat plateau extending to 10$U_{p}$
\cite{Paulus1994_2, Becker2002, Corkum1993}, which is originated from the elastic
scattering of electron when it revisits the parent ion.

In addition to the aforementioned features, an intensity-dependent
enhancement phenomenon in high-order ATI spectrum, the so-called
resonance-like enhancements (RLEs) was discovered in the rare gases Xenon
\cite{Hansch1997} and Argon \cite{Hertlein1997} two decades ago, triggering
great interest both experimentally and theoretically.
In 2003, Grasbon \textit{et al} \cite{Grasbon2003} discovered that RLEs are pronounced in
multi-cycle laser pulses and will be suppressed in few-cycle pulses.
Recently, the RLEs were also observed experimentally in the molecular system
including $H_{2}$ \cite{Cornaggia2010}, $N_{2}$ \cite{Quan2013}, formic acid
\cite{Wang2014}, the polyatomic molecules $C_{2}H_{4}$ and $C_{2}H_{6}$
\cite{Wang2016}.
RLEs are very sensitive to the laser intensity, and only a few
percent increases in the laser intensity will result in the ATI peaks on the
plateau being enhanced by up to an order of magnitude.
In the past two decades, great efforts have been made to study the underlying mechanism,
but no consensus has yet been reached.

RLEs are indeed the quantum effects of atoms or molecules in the intense
laser field, and cannot be described by the classical (semiclassical)
three-step model \cite{Corkum1993}, although it has successfully explained
the main features of ATI spectrum.
The theoretical calculations of the time-dependent Schr\"{o}dinger equation (TDSE)
in the single active electron approximation are in remarkable agreement with the experimental data,
indicating that high-order ATI can be explained by single electron dynamics \cite{Nandor1999}.
The interpretations based on the results of TDSE
\cite{Muller1998, Muller1999, Wassaf2003} and the Floquet approach
\cite{Wassaf2003, Wassaf2003_2, Potvliege2006, Potvliege2009} suggest that the
RLE structures are attributed to multiphoton resonance with laser-dressed
excited states, analogous to the Freeman resonances at low intensities
\cite{Freeman1987}.

The further theoretical researches based on the Strong Field Approximation
(SFA) obtain similar enhanced structures without considering
excited states \cite{Kopold1999, Paulus2001, Kopold2002, Popruzhenko2002,
Milosevic2007, Milosevic2008, Milosevic2010, Lai2013}.
The explanations of SFA argue that the kinetic energy of an ejected electron in
ATI is $\varepsilon_n = n\omega-I_p -U_p$, where $I_p$ is the ionization potential.
Hence, when $m\omega=I_p + U_p$, there are many electrons released with
near-zero kinetic energies that will be driven by the
multi-cycle laser field into recollisions with the parent ion many times,
and the constructive interference of a multitude of such trajectories will result
in the RLEs.

The semiclassical interpretations based on SFA grasp the main features that
RLEs result from the constructive interference of rescattering trajectories.
However, although SFA can give similar enhanced structures, the location and
amplitude of the enhanced structures are different from those of TDSE.
Additionally, Kopold \textit{et al} \cite{Kopold2002} have shown that SFA
can only get a partial qualitative result compared with the experiment.

Therefore, to better understand RLEs and clarify the contribution of excited
and continuum states to RLEs, a new fully quantum model is
established from TDSE without any excited states in this letter.
In the calculations of TDSE and model,
two types of enhanced structures, unimodal and multimodal structures, are observed.
Based on the calculations of the model, we know the RLEs are due to the constructive
interference of different momentum transfer channels caused by the
combination of the laser field and the ionic potential.
For the features of multi-channel interference, the Fano-like lineshapes are also discussed in
the end.

We start with the TDSE in the single active electron approximation
and the Hamiltonian within the length gauge reads
\begin{eqnarray}
H(t)
& = &-\frac{1}{2}\frac{\partial ^{2}}{\partial x^{2}}+V(x)+W(x, t)  \nonumber \\
& = &H_{a}+W(x, t)=H_{V}+V(x),
\label{Hamiltonian}
\end{eqnarray}%
where $V(x)=-1/\sqrt{x^{2}+a^{2}}$ represents the soft-core electron-nucleus
interaction \cite{Javanainen1988, Su1991}, with the soft-core parameter being $a^{2}=0.484$ to
match the first ionization energy of the helium atom. $W(x, t)=xE(t)$
denotes the atom-field interaction. $H_{a}$ and $H_{V}$ stand for the
atomic Hamiltonian and the Hamiltonian of a free electron in an external
field, respectively.

The wave functions of bound states (i.e., the functions whose corresponding
eigenvalues are negative among the eigenfunctions of $H_{a}$) and the Volkov
functions $
\psi _{p}(x,t)=(2\pi)^{-1/2} \exp\{i[p+A(t)]x-i
\int^{t}[p+A(\tau)]^{2}/2 d\tau\}$
can be used to construct an overcomplete basis of the Hilbert space.
Then, we expand the wave function by using the overcomplete basis,
\begin{equation}
\Phi (x,t)
= \sum_{n}c_{n}(t)\varphi _{n}(x,t)+\sum_{p}a_{p}(t)\psi _{p}(x,t),
\label{expansion}
\end{equation}%
in which $\varphi _{n}(x,t)=\varphi _{n}(x)e^{-iE_{n}t}$, while $\varphi
_{n}(x)$ is the wave function of $n$-th bound state of $H_{a}$ and $E_{n}$
is the corresponding eigenenergy.

Substituting the expansion $\eqref{expansion}$ into the Schr\"{o}dinger
equation, multiplying the conjugate of the basis functions $\psi^{\dagger} _{q} (x, t)$
and $\varphi^{\dagger} _{m} (x, t)$,
and assuming that the populations of all the excited states are negligible,
we can get
\begin{eqnarray}
i\frac{\partial}{\partial t} a_{q}(t)
&=&\sum\limits_{p}a_{p}(t)\left(V_{qp}-\Lambda _{qg}V_{gp}\right)+c_{g}(t)W_{qg},
\label{fullmodelA}\\
i\frac{\partial}{\partial t} c_{g}(t)
&=&\sum\limits_{p}a_{p}(t)V_{gp},
\label{fullmodelB}
\end{eqnarray}%
where the matrix elements of the electron-nucleus interaction are
$V_{qp}=\int\psi _{q}^{\dagger }(x,t)V(x)\psi_{p}(x,t)dx$
 and
$ V_{gp}=\int \varphi _{g}^{\dagger}(x,t)V(x)\psi _{p}(x,t)dx$,
the matrix elements of atom-field interaction are
$W_{qg}=\int \psi _{q}^{\dagger }(x,t)W(x,t)\varphi _{g}(x,t)dx$.
The term
$\Lambda _{qg}=\int \psi _{q}^{\dagger }(x,t)\varphi _{g}(x,t)dx$
is caused by the overcomplete basis.

Furthermore, the infinite summation of momentum should be truncated in
practice, so equations \eqref{fullmodelA} and \eqref{fullmodelB} are changed into
\begin{eqnarray}
i\frac{\partial}{\partial t} a_{q}(t)
&=&\sum\limits_{p}a_{p}(t)V_{qp}^{\ast }+c_{g}(t)W_{qg},
\label{modelA}\\
i\frac{\partial}{\partial t} c_{g}(t)
&=&\sum\limits_{p}a_{p}(t)V_{gp}^{\ast },
\label{modelB}
\end{eqnarray}%
where $V_{qp}^{\ast }=V_{qp}f(p,t)-\Lambda _{qg}V_{gp}g(p,t)$,
$V_{gp}^{\ast }=V_{gp}g(p,t)$.
The truncation functions $f(p,t)$ and $g(p,t)$ are
chosen as
\begin{eqnarray}
f(p,t)=
\begin{cases}
1, \qquad $if$~[p+A(t)]^2 \leq 2B~[B=3.17U_p], \\
\exp[-b(|p+A(t)|-\sqrt{2B})^2], ~ $otherwise$,
\end{cases} \\
g(p,t)=
\begin{cases}
1, ~$if$~[p+A(t)]^2 \leq 2C~[C=\min(I_p/2, U_p)], \\
\exp[-b(|p+A(t)|-\sqrt{2C})^2], \qquad $otherwise$,
\end{cases}
\end{eqnarray}
where we set the damping rate of the truncation functions $b=5$.
Here, the truncation function $f(p,t)$ is set by a
physical consideration that the maximum energy of the returned electron is
3.17$U_p$ \cite{Krause1992, Paulus1994}, and the truncation energy of $g(p,t)$
is set empirically \cite{Supplemental Material} by
comparing with the simulations of TDSE.
Consequently, we have derived a fully quantum model
(Eqs. \eqref{modelA} and \eqref{modelB}) from the TDSE without any excited states.
The details on the model derivation can be found in Sec. I of the Supplemental
Material \cite{Supplemental Material}.

We use the fourth-order Runge-Kutta algorithm and the split-operator method \cite{feim1982} to
numerically solve the model and TDSE, respectively.
The initial state, ground state, is obtained by applying the imaginary-time propagation.
A time step of 0.05 a.u. and 16384 grid points with a spatial step of 0.25 a.u. are used.

\begin{figure}[!htb]
\centering
\includegraphics[width=\linewidth]{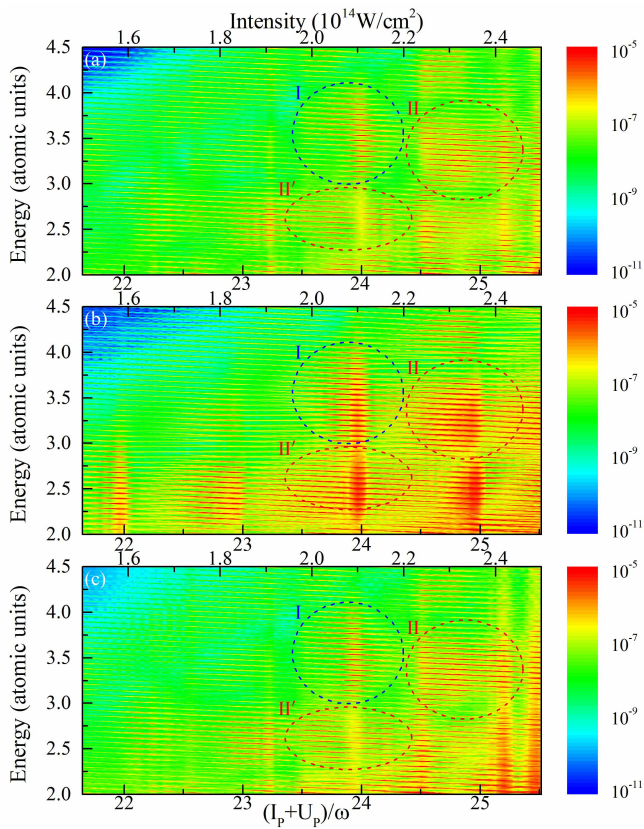}
\caption{ Photoelectron energy spectra calculated by (a) TDSE, (b) SFA,
and (c) model,
as a function of the laser intensity or the parameter $(I_p+U_p)/\protect\omega$.
Three islands I, II, II$'$ are marked.}
\label{fig1}
\end{figure}

In our calculations, we use 800 nm laser pulses with a total duration of 16
optical cycles, switched on and off linearly over 2 cycles.
To observe the RLEs, we scan the laser intensity from 1.5 to 2.5$\times 10^{14}\mathrm{W/cm^{2}}$
and plot the energy spectra in Fig. \ref{fig1}.
All the results calculated by TDSE, SFA \cite{Keldysh1965, Faisal1973, Reiss1980},
and our model show the pronounced RLEs.
The enhanced regions are irregularly located around some intensities
for the calculations of one-dimensional TDSE (Fig. \ref{fig1}(a)).
This phenomenon that the enhanced regions are not related to $(I_{p}+U_{p})/\omega$
can also be found in the energy spectra of Xenon calculated by three-dimensional TDSE \cite{Li2018}.
However, the enhancements are always pronounced at integer values of $(I_{p}+U_{p})/\omega $
both in the calculations of one-dimensional SFA (Fig. \ref{fig1}(b)) and three-dimensional SFA \cite{Milosevic2007}.

The calculations of model (Fig. \ref{fig1}(c)) reproduce the RLEs similar to those of TDSE.
The RLEs depend on the intensity and energy,
leading to the islands in the energy spectra.
There are two types of patterns for islands in the calculations of TDSE and model.
The enhanced structures in the island I are pronounced in the center of the intensity of the island,
while the enhancements in the islands II and II$'$ are distributed over a wide range of the intensity.

\begin{figure}[!htb]
\centering
\includegraphics[width=\linewidth]{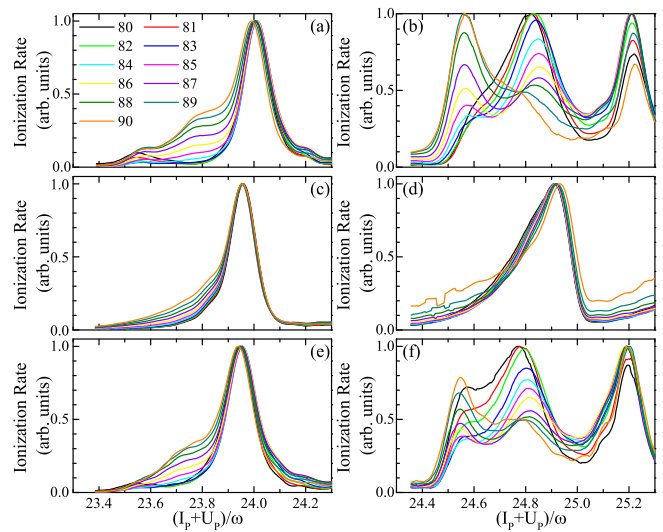}
\caption{ Left panel: Normalized differential ionization rate calculated by (a) TDSE, (c) SFA,
and (e) model as a function of $(I_p+U_p)/\protect\omega$ for island I.
Right panel: Normalized differential ionization rate calculated by (b) TDSE, (d) SFA,
and (f) model as a function of $(I_p+U_p)/\protect\omega$ for island II.
All the orders of ATI peaks, i.e., photon numbers absorbed by electrons, for different lines are labeled in (a).}
\label{fig2}
\end{figure}

To observe the elaborate structures, we plot the ionization rate as a function of $(I_p+U_p)/\protect\omega$ for different ATI peaks in Fig. \ref{fig2}.
The rate is the normalization of the summation of the probability of energy range from
$\varepsilon_n-\omega/2$ to $\varepsilon_n+\omega/2$.
The orders of the ATI peaks in each island can be seen in Sec. III of the Supplemental Material \cite{Supplemental Material}.
As plotted in Fig. \ref{fig2}, the enhanced structures of different ATI peaks in each island calculated by each method are qualitatively consistent.
The results of TDSE show there are two types of enhanced structures within one order difference of $(I_p+U_p)/\protect\omega$.
The structures in the island I (Fig. \ref{fig2}(a)) are unimodal and the structures in the island II (Fig. \ref{fig2}(b)) are multimodal.
On the contrary, the calculations of SFA (Figs. \ref{fig2}(c) and \ref{fig2}(d)) always exhibit unimodal structures.
The calculations of model are qualitatively consistent with the results of TDSE.
The enhanced structures in the islands I (Fig. \ref{fig2}(e)) and II (Fig. \ref{fig2}(f))
are unimodal and multimodal, respectively.
Besides, the enhanced structures in the island II$'$ calculated by TDSE, SFA, and model are qualitatively consistent with those in the island II (see Sec. III of the Supplemental Material \cite{Supplemental Material}).
The agreement between TDSE and model indicates
that the excited states are not fundamental to the RLEs because the model doesn't contain any
excited states.
The structures calculated by SFA and TDSE or model have an evident difference, indicating that the
underlying mechanisms of RLEs should be further studied.

\begin{figure}[!htb]
\centering
\includegraphics[width=\linewidth]{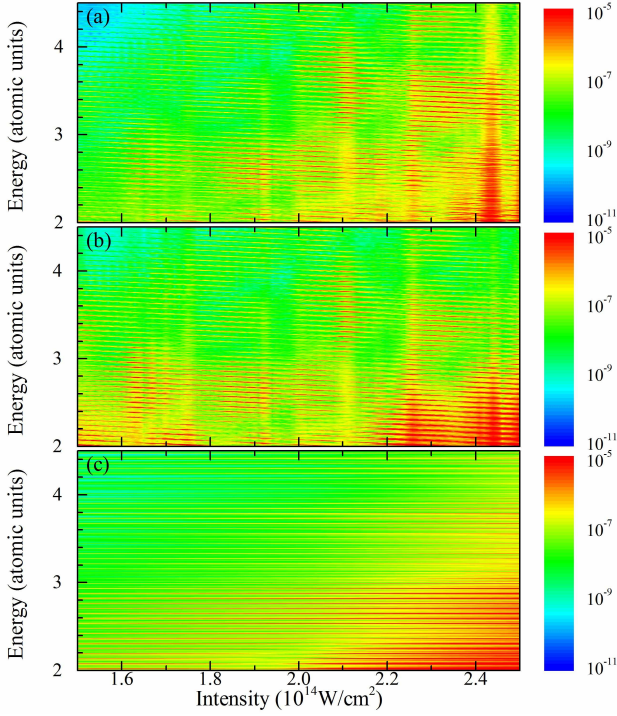}
\caption{ Photoelectron energy spectra calculated by (a) Eq. \eqref{modelNew},
(b) Eq. \eqref{modelNew} with the intensity of $W_{pg}$ is kept as 2$\times 10^{14} \mathrm{W/cm^{2}}$,
and (c) Eq. \eqref{modelNew} with the intensity of $V_{qp}^{\ast }$ is kept as 2$\times 10^{14} \mathrm{W/cm^{2}}$. }
\label{fig3}
\end{figure}

To clarify the mechanisms of RLEs, we analyze the three processes in the model
(also can see Sec. IV of the Supplemental Material \cite{Supplemental Material}).
The first process is $W_{pg}$, the direct ionization process
from the ground state. Another is $V_{qp}^{\ast }$, the transfer process
between the continuum states. The third is $V_{gp}^{\ast }$, the recombination
of the electron from the continuum state into the ground state.
After assuming $c_{g}(t)=1$, we can write
\begin{equation}
i\frac{\partial}{\partial t} a_{q}(t)
= \sum\limits_{p}a_{p}(t)V_{qp}^{\ast }+W_{qg}.
\label{modelNew}
\end{equation}
As plotted in Fig. \ref{fig3}(a), the energy spectra calculated by Eq. \eqref{modelNew} also reproduce the enhanced structures,
which are similar to the results of TDSE and model.
It reveals that the dynamics of the ground state is not the key factor in the RLEs.

Next, we separately fix the laser intensities of one of the two processes, namely $W_{pg}$ and $V_{qp}^{\ast }$,
and investigate the intensity-dependent effect of the other process.
In Fig. \ref{fig3}(b), we fix
the laser intensity of $W_{pg}$ to 2$\times 10^{14}\mathrm{W/cm^{2}}$ and
scan the intensity of the transfer process.
There are still resonance pattern in the energy spectra, indicating that the direct ionization process
is not the key factor in the RLEs.
Besides, the electron liberated with finite kinetic energy (even greater than one photon energy)
during the direct ionization process contributes significantly to RLEs
(see Sec. V of the Supplemental Material \cite{Supplemental Material}).
On the contrary, in Fig. \ref{fig3}(c), we keep the laser intensity of $V_{qp}^{\ast }$
at 2$\times 10^{14}\mathrm{W/cm^{2}}$ and scan the intensity of $W_{pg}$.
The enhanced structures disappear in the energy spectra, indicating that the RLEs are mainly
caused by the interference of momentum transfer processes.
Different momenta may transfer to the same final momentum, and each transfer channel
has a different phase.
As the intensity increases, the phase changes accordingly.
For some intensities, the constructive interferences occur and lead to the RLEs.

\begin{figure}[!htb]
\centering
\includegraphics[width=\linewidth]{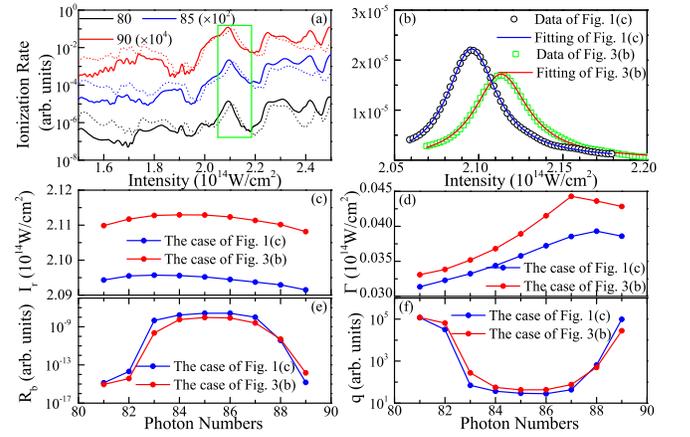}
\caption{ (a) Differential ionization rate as a function of the laser
intensity for the case of Fig. \protect\ref{fig1}(c) (solid lines) and Fig.
\protect\ref{fig3}(b) (dotted lines) with photon numbers absorbed by
electrons $n=80, 85, 90$, where the results of different $n$ are
multiplied by different multiples to facilitate separation.
(b) The data and the fitting result as a function of the laser
intensity for photon numbers $n=85$.
The fitting parameters (c) $I_r$, (d) $\Gamma$, (e)$ R_b$, and (f) $q$
as a function of photon numbers. }
\label{fig4}
\end{figure}

To further explore the effect of the direct ionization on the enhanced structures,
we plot the differential ionization rate,
which is the summation of the probability of energy range from
$\varepsilon_n-\omega/2$ to $\varepsilon_n+\omega/2$,
as a function of the laser intensity in Fig. \ref{fig4}(a).
Here the solid lines present the case of Fig. \ref{fig1}(c)
and the dotted lines present the case of Fig. \ref{fig3}(b).
Their oscillating structures varying with the intensity are similar.
The difference in amplitudes reveals that direct ionization is not negligible in the ionization rate.
For the scanned intensity less than the fixed intensity (2 $\times 10^{14}\mathrm{W/cm^{2}}$),
the ionization rate is overestimated.
When the scanned intensity is greater than the fixed intensity,
the rate is underestimated in most cases.

To our knowledge, such resonant structure caused by the multi-channel interference may have the shape of Fano
resonance and its lineshape can be fitted by the Fano formula
\cite{Fano1961}
\begin{equation}
R
= R_b\frac{(\epsilon+q)^2}{\epsilon^2+1},
\label{fano}
\end{equation}
where $\epsilon = 2(\varepsilon-\varepsilon_r)/\Gamma_\varepsilon = 2(I_0-I_r)/\Gamma $
is the reduced energy with the resonance width $\Gamma = 4\omega^2\Gamma_\varepsilon$
by using the relation $\varepsilon = n\omega-I_p-U_p$,
 $R_b$ is the smoothly varying background parameter,
$q$ is the Fano asymmetry parameter, $I_0 = E_0^2$, and
$I_r$ is the resonant position.
We choose the isolated unimodal structures marked in Fig. \ref{fig4}(a)
(also can be seen in Fig. \ref{fig2}(e))
to study the Fano resonance and the contribution of direct ionization.
We use the Fano formula \eqref{fano} to fit the isolated unimodal structures.
All the coefficients of determination R-squared are greater than 0.995,
indicating that the fitting results can well reproduce the calculated curves.
Taking the photon number $n=85$ as a sample, we plot the calculated and fitting results in Fig. \ref{fig4}(b).
The fitting parameters $I_r$, $\Gamma$, $R_b$, and $q$ are displayed in Figs. \ref{fig4}(c)$-$\ref{fig4}(f).
The four fitting parameters varying with the photon numbers in the two cases are closely similar.
As photon numbers increase, the resonant position $I_r$ varies slightly and the resonance width $\Gamma$ increases slowly between
3 and 4.5$\times10^{12}\mathrm{W/cm^{2}}$.
The different values in the two cases shows that the direct ionization affects the resonant position and resonance width.
Figures \ref{fig4}(e) and \ref{fig4}(f) show that the trends of $R_b$ and $q$ in both cases are basically the same,
where they vary slightly around $n=85$ but change rapidly at $n=82$ and 88.
This means that for the selected region of RLEs, the ATI peak near the 85th order is the center of resonance
and the ATI peaks around the 82nd and 88th orders are critical regions.
Hence, we have studied the enhanced structures from two dimensions of laser intensity and ATI peak by using the Fano formula.
There are resonance centers and resonance widths in both dimensions.
Moreover, the direct ionization affects the position, width, and magnitude of the enhanced structures,
although it isn't the key factor in the RLEs.

Besides, the fitting structures tend to the Lorentzian profiles $R=R'_b(\Gamma/4)/[(I_0-I_r)^2+(\Gamma/2)^2]$, due to the parameter $q$ is large.
For the Lorentzian profiles, the line width depends on the lifetime of the energy levels and the collision broadening \cite{Loudon2000}.
In the calculation, we don't consider any decay of all the states because the excited states are not taken into account in the model.
Hence, the line width is caused by the collision broadening and the average collision interval $\tau_{coll}=1/(\pi\Gamma_\varepsilon)$ between electron and ion is in the range of $T/2.29$ to $T/3.43$, where $T$ is the laser period.

In summary, we have derived a new quantum model without excited states from TDSE.
We have observed that the differential ionization rate of each ATI peak as a function of the laser intensity
has two types of enhanced structures in the calculations of TDSE and our model.
One is the same unimodal structure at the integer values of $(I_{p}+U_{p})/\omega $ as the calculation of SFA,
the other is a multimodal structure within one order difference of $(I_{p}+U_{p})/\omega $.
Based on the calculations of our model, we show the constructive interference of different
momentum transfer channels, rather than the excited states, lead to the RLEs.
Besides, the direct ionization affects the position, width, and magnitude of the enhanced structures.
Finally, the Fano-like lineshapes are studied to explore the dependence of RLEs
on the laser intensity and the ATI peak,
and to confirm the contributions of the transfer and the direct ionization processes.
We hope that the abundant structures, such as the unimodal or multimodal enhanced structures
and the Fano lineshapes, will be observed in experiments in the future.

This work was supported by National Natural Science Foundation of China
(Grant No. 11725417, 11575027), NSAF (Grant No. U1730449), and Science
Challenge Project (Grant No. 2018005).

\end{document}